\documentclass[12pt]{article}
\begin{document}

\title{\bf Energy-Momentum Distribution of Non-Static Plane Symmetric
Spacetimes in GR and TPT}

\author{M. Jamil Amir, \thanks {mjamil.dgk@gmail.com} Sarfraz Ali
\thanks{sarfraz270@yahoo.com} and Tariq Ismaeel \thanks {Tariqismaeel@hotemail.com}
\\Department of Mathematics, University of Sargodha,\\Pakistan,\\
Department of Mathematics, University of Education, Okara,\\Pakistan,\\
Department of Mathematics, GC University Lahore,\\Pakistan. \\
}

\date{}

\maketitle

\begin{abstract}
This paper is devoted to explore the energy-momentum of non-static
plane symmetric spacetimes in the context of General Relativity
and teleparallel theory of gravity. For this purpose, we use four
prescriptions, namely, Einstein, Landau-Lifshitz, Bergmann-Thomson
and M{\o}ller in both theories. It is shown that the results for
the first three prescriptions turn out to be same in both the
theories but different for last prescription. It is mentioning
here that our results coincide with the results obtained by Sharif
and kanwal [1] for Bell-Szekeres metric under certain choice of
the metric functions.
\end{abstract}


\section{Introduction}

Among all the available theories of gravitation in literature, the
theory of General Relativity (GR) is considered as a standard
theory of gravitation due to the fact that many physical aspects
of nature have been verified experimentally in this theory.
However, the problem of localization of energy and momentum in GR,
is still most controversial [2]. A number of scientists have
attempted to resolve this issue and gave their own definitions. As
a pioneer, Einstein [3] gave an energy-momentum prescription for
the localization of energy and momentum. Following him, many well
known scientists like, Landau-Lifshitz [4], M{\o}ller [5],
Bergmann-Thomson [6], Tolman [7] and Weinberg [8] gave their own
energy-momentum prescriptions. To explore energy, the use of
Cartesian coordinates are necessary for these prescriptions except
M{\o}ller's prescription, which is independent of the coordinate
system. Misner et, al. [2] proved that the energy can be localized
in spherical coordinate system. After a short time, Cooperstok and
Sarracino [9] proved that if the energy can be localized in
spherical system then it can be localized in any other coordinate
system. Virbhadra and his collaborators [10-12] explored the
energy-momentum distribution of several spacetimes, such as,
Kerr-Newmann, Kerr-Schild classes, Einstein-Rosen, Vaidya and
Bonnor-Vaidya spacetimes. They showed that different
energy-momentum prescriptions provide the same results which agree
with those obtained by Penrose [13] and Tod [14] in the framework
of quasi-local mass. Einstein [15] used the notion of tetrad field
to unify gravitation and electromagnetism but he was not succeeded
in his purpose. Hayashi and Nakano [16] formulated the tetrad
theory of gravitation, which is known as teleparallel theory (TPT)
of gravity or new General Relativity. This theory based on non-
trivial tetrad fields and is defined on Weitzenb\"{o}ck [17]
geometry. The curvature tensor of Weitzenb\"{o}ck  connection
vanishes identically but torsion remains non zero. In the frame
work of TPT, gravitation is attributed to torsion [18] which plays
the role of force while it geometrizes the underlying spacetime in
the case of GR.

A number of scientists [19]  hoped that the problem of
localization of energy might be resolved in the frame work of TPT
of gravity and results may coincide with those already existing in
GR. Vargas [20] showed that total energy of the closed
Friedmann-Robertson-Walker (FRW) universe is zero by using
teleparallel (TP) version of Einstein and Landau-Lifshitz
prescriptions. The results obtained by Vargas coincide with those
found by Rozen [21]. This opened the task for many authors who
explored the energy momentum distribution of many spacetimes by
using the TP version of different prescriptions. These
prescriptions yield same results for some spacetimes and different
in the case of others. Pereira at. al, [22] obtained the TP
version of Schwarzschild and stationary axisymmetric Kerr
solutions. Sharif and Jamil [23] found the TP versions of
Friedmann models and Lewis-Papapetrou spacetimes and obtained
interesting results. They [24] also explored the energy momentum
distribution of the Lewis-Papapetrou spacetime by using the TP
version of M{\o}llar prescription. They [25, 26] extended this
work to stationary axisymmetric solutions of Einstein-Maxwell
field equations and the Levi-Civita vacuum solutions. They [27,
28, 29] also explored the energy-momentum distribution of static
axially symmetric, Friedmann models and the spatially homogenous
rotating spacetimes by using different prescriptions in the
context of TPT. Sharif and Kanwal [1] explored the energy-momentum
distribution of the Bell-Szekeres metric in GR and TPT and showed
that the four prescriptions ELLBTM yield same results in both the
theories . Recently, Sharif and Sumaira [30] used Hamiltonian
formulation of TEGR to explore the energy-momentum of Non-Vacuum
Spacetimes  and found consistent results.

The scheme of paper is as follows: in section $2$ an over view of
TPT is given. The different energy-momentum prescriptions in both
GR and TPT are given in section $3$. Section $4$ is devoted to
explore the energy-momentum distribution of non-static plane
symmetric spacetimes in GR. The section $5$, contains the
energy-momentum distribution of non-static plane symmetric
spacetimes in TPT. The last section furnishes the summery and
discussion of the results obtained.

\section{An Overview of the Teleparallel Theory}

TPT is based on Weitzenb$\ddot{o}$ck connection given as [31]
\begin{eqnarray}
{\Gamma^\theta}_{\mu\nu}={{h_a}^\theta}\partial_\nu{h^a}_\mu,
\end{eqnarray}
where ${h_a}^\nu$ is a non-trivial tetrad. Its inverse field is
denoted by ${h^a}_\mu$ and satisfy the relations
\begin{eqnarray}
{h^a}_\mu{h_a}^\nu={\delta_\mu}^\nu; \quad\
{h^a}_\mu{h_b}^\mu={\delta^a}_b.
\end{eqnarray}
Here the Latin alphabet $(a,b,c,...=0,1,2,3)$ are used to denote
tangent space indices and the Greek alphabet
$(\mu,\nu,\rho,...=0,1,2,3)$ to denote spacetime indices. The
Riemannian metric in TPT arises as a product [31] of the tetrad
field given by
\begin{equation}
g_{\mu\nu}=\eta_{ab}{h^a}_\mu{h^b}_\nu,
\end{equation}
where $\eta_{ab}$ is the Minkowski metric. For the
Weitzenb$\ddot{o}$ck spacetime, the torsion is defined as [31]
\begin{equation}
{T^\theta}_{\mu\nu}={\Gamma^\theta}_{\nu\mu}-{\Gamma^\theta}_{\mu\nu}
\end{equation}
which is antisymmetric w.r.t. its last two indices. Due to the
requirement of absolute parallelism, the curvature of the
Weitzenb$\ddot{o}$ck connection vanishes identically. The
Weitzenb$\ddot{o}$ck connection also satisfies the relation
\begin{equation}
{{\Gamma^{0}}^\theta}_{\mu\nu}={\Gamma^\theta}_{\mu\nu}
-{K^{\theta}}_{\mu\nu},
\end{equation}
where
\begin{equation}
{K^\theta}_{\mu\nu}=\frac{1}{2}[{{T_\mu}^\theta}_\nu+{{T_\nu}^
\theta}_\mu-{T^\theta}_{\mu\nu}]
\end{equation}
is the {\bf contortion tensor} and ${{\Gamma^{0}}^\theta}_{\mu\nu}
$ are the Christoffel symbols in GR.

\section{Energy-Momentum Complexes}

To explore the energy-momentum distribution of a given spacetime
in the framework of GR, different approaches have been used by the
different scientists. To derive an energy-momentum complex for the
localization of energy and momentum is one of these approaches.
The Einstein, Landau-Lifshitz, Bergmann-Thomson and M{\o}ller
complexes in both GR and TPT are given as:

\subsection{Energy-Momentum Complexes in GR}

For \textbf{Einstein }prescription, the energy-momentum density
components are given by [3]
\begin{equation}
\Theta^b_a=\frac{1}{16 \pi}{H^{bc}}_{a,c},
\end{equation}
where ${H^{bc}}_{a}$ is a function of metric tensor and its first
order derivatives given as
\begin{equation}
{H^{bc}}_{a}=\frac{g_{ad}}{\sqrt{-g}}[-g(g^{bd}g^{ce}-g^{cd}g^{be})],_{e}.
\end{equation}
Here $\Theta^{0}_{0}$ stands for energy density and
$\Theta^{0}_{i}(i=1,2,3)$ are the momentum density components and
$\Theta^{i}_{0}$ are the current density components. The momentum
four-vector is
\begin{equation}
P_a ={\int}_V {\Theta^{0}_{a} dx^{1}dx^{2}dx^{3}}.
\end{equation}
and the energy of the physical system is
\begin{equation}
P_0 ={\int}_V {\Theta^{0}_{0} dx^{1}dx^{2}dx^{3}}.
\end{equation}
It is mentioned here that these calculations are restricted to be
done in Cartesian coordinates only to obtain physical results.

For \textbf{Landau-Lifshitz} prescription, the energy-momentum
density components are given as [4]
\begin{equation}
L^{ab}= \frac{1}{16\pi} l^{abcd},_{cd},
\end{equation}
where
\begin{equation}
l^{abcd}= (-g)(g^{ab}g^{cd}-g^{ac}g^{bd}).
\end{equation}
The quantity $L^{00}$ gives the energy density component of whole
system and $L^{i0}~(i=1,2,3)$ represents the momentum density
components.

For \textbf{Bergmann-Thomson} prescription, the energy-momentum
density components are given by [6]
\begin{equation}
B^{ab}=\frac{1}{16\pi}M^{abc},_{c},
\end{equation}
where
\begin{equation}
M^{abc}=g^{ad}{V^{bc}}_{d}
\end{equation}
and
\begin{equation}
{V^{bc}}_{d}=\frac{g_{de}}{\sqrt{-g}}[-g(g^{be}g^{cf}-g^{ce}g^{bf})]_,{f}.
\end{equation}
The quantity $B^{00}$ represents energy density of the whole
system and $B^{i0}(i=1,2,3)$ represents the momentum density
components.

Einstein, Landau-Lifshitz and Bergmann-Thomson energy-momentum
prescriptions are coordinate dependent while \textbf{M{\o}ller}
introduced another energy-momentum pseudo-tensor $M^{b}_{a}$ which
is coordinate independent, given as
\begin{equation}
M^{b}_{a}=\frac{1}{8\pi}{K^{bc}}_{a,c},
\end{equation}
where
\begin{equation}
{K^{bc}}_{a}=\sqrt{-g}(g_{ad,e}-g_{ae,d})g^{be}g^{cd}.
\end{equation}
Clearly, ${K^{bc}}_{a}$ is antisymmetric w.r.t. its upper indices.
$M^{0}_{0}$ is the energy density and $M^{0}_{i}(i=1,2,3)$ are the
momentum density components and $M^{i}_{0}(i=1,2,3)$ are the
components of current density. The momentum four-vector is given
by
\begin{equation}
p_{a}=\int\int_{V}\int M^{0}_{a}dx^{1}dx^{2}dx^{3},
\end{equation}
where $p_0$ gives the energy and $p_i (i=1,2,3)$ give the
momentum. Using Gauss's theorem, the total energy-momentum
components may be given in the form of surface integral as
\begin{equation}
p_{a}=\frac{1}{8\pi}\int_{S}\int K^{0c}_{a}\bf{n_{c}}.dS,
\end{equation}
where $\bf{n_{c}}$ is the outward unit normal vector over an
infinitesimal surface element \textbf{dS}.

\subsection{Energy-Momentum Complexes in TPT}

It was noticed that tetrad description of the gravitational field
allows more satisfactory treatment of the gravitational
energy-momentum. The Gauge field Lagrangian is given as
\begin{equation}
L=\frac{h}{16G\pi}[\frac{1}{4}{F^{a}}_{\mu\nu}{F^{b}}_{\theta\rho}
g^{\mu\theta}{N_{ab}}^{\nu\rho}],
\end{equation}
where $h=det({h^{a}}_{\mu})$, G is the gravitational constant and
${F^{a}}_{\mu\nu}$ is field strength. In the presence of tetrad
field, algebra and spacetime induces can be interchange and
consequently it appears mixed up in the Lagrangian. It means that
\begin{equation}
{N_{ab}}^{\nu\rho}=\eta_{ab}g^{\nu\rho}=\eta_{ab}h^{\nu}_{c}h^{c\rho}
\end{equation}
must now include all cyclic permutations of a, b and c. A simple
calculation shows that
\begin{equation}
{N_{ab}}^{\nu\rho}=\eta_{ab}h^{\nu}_{c}h^{c\rho}+2
{h_{a}}^{\rho}{h_{b}}^{\nu}-4{h_{a}}^{\nu}{h_{b}}^{\rho}.
\end{equation}
Substituting Eq.(22) in Eq.(20), we get
\begin{equation}
L=\frac{h}{16G\pi}{F^{a}}_{\mu\nu}{F^{b}}_{\theta\rho}g^{\mu\theta}
[\frac{1}{4}{h^{\nu}}_{c}h^{c\rho}\eta_{ab}+\frac{1}{2}{h_{a}}^{\rho}
{h_{b}}^{\nu}-{h_{a}}^{\nu}{h_{b}}^{\rho}].
\end{equation}
Using the value of field strength
${F^{a}}_{\mu\nu}=c^{2}{h^{a}}_{\rho}{T^{\rho}}_{\mu\nu}$ in
Eq.(23), we have
\begin{eqnarray*}
L&=&\frac{hc^{4}}{16G\pi}[\frac{1}{4}{T^{\rho}}_{\mu\nu}{T_{\rho}}^{\mu\nu}
+\frac{1}{2}{T^{\rho}}_{\mu\nu}{T^{\nu\mu}}_{\rho}-{T_{\rho\mu}}^{\rho}
{T^{\nu\mu}}_{\nu}],\\
&=&\frac{hc^{4}}{16G\pi}
{S_{\rho}}^{\mu\nu}{T^{\rho}}_{\mu\nu},\\
L&=&\frac{hc^{4}}{16G\pi} S^{\rho\mu\nu}T_{\rho\mu\nu},\nonumber
\end{eqnarray*}
where
\begin{equation}
S^{\rho\mu\nu}=\frac{1}{4}[T^{\rho\mu\nu}+T^{\mu\rho\nu}+T^{\nu\rho\mu}]
+\frac{1}{2}[g^{\rho\mu\nu}{T^{\theta\nu}}_{\theta}-g^{\rho\nu}
{T^{\theta\mu}}_{\theta}]
\end{equation}
is a tensor written in terms of the Weitzenb\"{o}ck connection.
Now, the Freud's superpotential is defined as
\begin{equation}
{U_{\rho}}^{\mu\nu}=h{S_{\rho}}^{\mu\nu}.
\end{equation}
Vargas [20] gave TP version of Einstein, Bergmann-Thomson and
Landau-Lifshitz prescriptions by using this superpotential as
\begin{eqnarray}
hE_\nu^\mu&=&\frac{1}{4\pi}\partial_\lambda({U_\nu}^{\mu\lambda}),\\
hL^{\mu\nu}&=&\frac{1}{4\pi}\partial_\lambda(hg^{\mu\beta}{U_\beta}
^{\nu\lambda}),\\
hB^{\mu\nu}&=&\frac{1}{4\pi}\partial_\lambda({g^{\mu\beta}U_\beta}
^{\nu\lambda}).
\end{eqnarray}
The four-vector momentum for these complexes are given in the
following relations.
\begin{equation}
p^{E}_{\mu}=\int_{\Sigma}hE^{0}_{\mu}dxdydz,
\end{equation}
\begin{equation}
p^{B}_{\mu}=\int_{\Sigma}hB^{0}_{\mu}dxdydz,
\end{equation}
\begin{equation}
p^{L}_{\mu}=\int_{\Sigma}hL^{0}_{\mu}dxdydz,
\end{equation}
where $p_{0}$ and $p_{i}(i=1,2,3)$ represent energy and momentum
components respectively. In Eqs. (29) to (31), the integration is
taken over the hypersurface $\Sigma$ obtained by taking $t=$
constant.

Now, we discuss M{\o}ller energy-momentum complex in the context
of TPT. Mikhail et al. [19] defined the superpotential (which is
antisymmetric in its last two indices) of the M{\o}ller tetrad
theory as
\begin{equation}
{U_{\mu}}^{\nu\beta}=\frac{\sqrt{-g}}{2\kappa}P^{\tau\nu\beta}_{\chi\rho\sigma}
[V^{\rho}g^{\sigma\chi}g_{\mu\tau}-\lambda
g_{\tau\mu}K^{\chi\rho\sigma}-g_{\tau\mu}(1-2\lambda)K^{\sigma\rho\chi}],
\end{equation}
where
\begin{equation}
P^{\tau\nu\beta}_{\chi\rho\sigma}=\delta^{\tau}_{\chi}g^{\nu\beta}_{\rho\sigma}
+\delta^{\tau}_{\rho}g^{\nu\beta}_{\sigma\chi}-\delta^{\tau}_{\sigma}
g^{\nu\beta}_{\chi\rho}
\end{equation}
and $g^{\nu\beta}_{\rho\sigma}$ is a tensor quantity defined as
\begin{equation}
g^{\nu\beta}_{\rho\sigma}=\delta^{\nu}_{\rho}\delta^{\beta}_{\sigma}-
\delta^{\nu}_{\sigma}\delta^{\beta}_{\rho}.
\end{equation}
Here $K^{\sigma\rho\chi}$ is a contorsion tensor, $g$ is the
determinant of the metric tensor, $\kappa$ is the coupling
constant and $V^{\mu}$ is the basis vector field, which is given
by
\begin{equation}
V_{\mu}={T^{\nu}}_{\nu\mu}.
\end{equation}
In TPT, the M{\o}ller's energy-momentum density is then defined as
\begin{equation}
\Xi^{\nu}_{\mu}={{U_{\mu}}^{\nu\rho}}_{,\rho},
\end{equation}
where comma denotes ordinary differentiation. The energy E contained
in a sphere of radius R is expressed by the volume integral as
\begin{equation}
E(R)=\int_{r=R}{\Xi^{0}_{0}}~dxdydz,
\end{equation}
and the spatial momentum $p_{i}, (i=1,2,3)$ is given by
\begin{equation}
p_{i}(R)=\int_{r=R} \Xi^{0}_{i}~dxdydz.
\end{equation}

\section{Energy-Momentum Distribution in GR}

In this section, we explore  the energy-momentum distribution of
non-static plane symmetric spacetimes by using four different
prescriptions of GR. The line element representing non-static
plane symmetric spacetimes is given by [33]
\begin{equation}
ds^{2}=e^{2\nu(t,x)}dt^{2}-e^{2\mu(t,x)}dx^{2}-e^{2\lambda(t,x)}(dy^{2}+dz^{2}),
\end{equation}
Making use of Eq.(39) in Eq.(8), we get the following
non-vanishing components of ${H^{bc}}_{a}$
\begin{eqnarray}
{H^{01}}_{0}=-{H^{10}}_{0}&=&-4e^{\nu-\mu+2\lambda}\lambda_{x},\nonumber\\
{H^{01}}_{1}=-{H^{10}}_{1}&=&-4e^{\mu-\nu+2\lambda}\lambda_{t},\nonumber\\
{H^{02}}_{2}=-{H^{20}}_{2}&=&-2e^{\nu-\mu+2\lambda}(\lambda_{t}+\mu_{t}),\nonumber\\
{H^{03}}_{3}=-{H^{30}}_{3}&=&2e^{\mu-\nu+2\lambda}(\lambda_{t}+\mu_{t}),\nonumber\\
{H^{12}}_{2}=-{H^{21}}_{2}&=&2e^{\nu-\mu+2\lambda}(\nu_{x}
+\lambda_{x})={H^{13}}_{3}=-{H^{31}}_3.
\end{eqnarray}
Substituting Eq.(40) in Eq.(7), the non-zero energy-momentum
density components of \textbf{Einstein's} prescription turn out to
be
\begin{eqnarray}
\Theta^{00}&=&-\frac{1}{4\pi}[\lambda_{x}(2\lambda_x-\mu_{x}
+\nu_{x})+\lambda_{xx}]~e^{2\lambda-\nu-\mu},\\
\Theta^{10}&=&\frac{1}{4\pi}[\lambda_{t}(2\lambda_x+\mu_{x}
-\nu_{x})+\lambda_{tx}]~e^{2\lambda-\nu-\mu}.
\end{eqnarray}
Now, we substitute Eq.(39) in Eq.(12) and obtain the following
non-vanishing components of $l^{abcd}$ as
\begin{eqnarray}
l^{0101}=-l^{0110}=-l^{1001}=l^{1010}&=&e^{4\lambda},\nonumber\\
l^{1313}=-l^{1331}=-l^{3131}=l^{3113}&=&-e^{2\nu+2\lambda},\nonumber\\
l^{0202}=-l^{0220}=-l^{2002}=l^{2020}&=&-e^{2\mu+2\lambda},\nonumber\\
l^{0303}=-l^{0330}=-l^{3003}=l^{3030}&=&e^{2\mu+2\lambda},\nonumber\\
l^{1212}=-l^{1221}=-l^{2112}=l^{2121}&=&e^{2\nu+2\lambda},\nonumber\\
l^{2323}=-l^{2332}=-l^{3223}=l^{3232}&=&-e^{2\nu+2\mu}.
\end{eqnarray}
Making use of Eq.(43) in Eq.(11) yields the non-zero
energy-momentum density components of \textbf{Landau-Lifshitz's}
prescription as
\begin{eqnarray}
L^{00}&=&-\frac{1}{4\pi}e^{4\lambda}(4\lambda_{x}^{2}+\lambda_{xx}),\\
L^{10}&=&\frac{1}{4\pi}e^{4\lambda}(4\lambda_{x}\lambda_{t}+\lambda_{tx}).
\end{eqnarray}
Using Eq.(39) in Eq.(15), we get the following non-vanishing
components of ${V^{ab}}_{c}$ as
\begin{eqnarray}
{V^{10}}_{0}&=&-{V^{01}}_{0}=4e^{2\lambda+\nu-\mu}\lambda_{x},\nonumber\\
{V^{10}}_{1}&=&-{V^{01}}_{1}=4e^{2\lambda-\nu+\mu}\lambda_{t},\nonumber\\
{V^{20}}_{2}&=&-{V^{02}}_{2}=2e^{2\lambda-\nu+\mu}(\lambda_{t}+\mu_{t}),\nonumber\\
{V^{30}}_{3}&=&-{V^{03}}_{3}=2e^{2\lambda-\nu+\mu}(\lambda_{t}+\mu_{t}),\nonumber\\
{V^{31}}_{3}&=&-{V^{13}}_{3}=2e^{2\lambda+\nu-\mu}(\lambda_{x}+\nu_{x}).
\end{eqnarray}
Substituting the values from Eq.(46) in Eq.(14) and then in
Eq.(13), the non-zero energy-momentum density components of
\textbf{Bergmann-Thomson's} prescription turn out to be
\begin{eqnarray}
B^{00}&=&-\frac{1}{4\pi}[\lambda_{x}(2\lambda_{x}-\mu_{x}
-\nu_{x})+\lambda_{xx}]~e^{2\lambda-\nu-\mu}\\
B^{10}&=&\frac{1}{4\pi}[\lambda_{t}(2\lambda_{x}-\mu_{x}-\nu_{x})
+\lambda_{tx}]~e^{2\lambda-\nu-\mu}
\end{eqnarray}
The non-vanishing components of ${K^{ab}}_{c}$ are obtained by
using Eq.(39) in Eq.(17) as
\begin{eqnarray}
{K^{01}}_{0}&=&-{K^{10}}_{0}=2e^{\nu-\mu+2\lambda}\nu_{x},\nonumber\\
{K^{01}}_{1}&=&-{K^{10}}_{1}=2e^{\mu-\nu+2\lambda}\mu_{t},\nonumber\\
{K^{02}}_{2}&=&-{K^{20}}_{2}=2e^{\mu-\nu+2\lambda}\lambda_{t},\nonumber\\
{K^{03}}_{3}&=&-{K^{30}}_{3}=2e^{\mu-\nu+2\lambda}\lambda_{t}.
\end{eqnarray}
In view of Eq.(49), the non-zero energy-momentum density
components of \textbf{M{\o}ller's} prescription in contravariant
form are obtained from Eq.(16) as
\begin{eqnarray}
M^{00}&=&\frac{1}{4\pi}[\nu_{x}(2\lambda_{x}-\mu_{x}+\nu_{x}
)+\nu_{xx}]~e^{2\lambda-\nu-\mu},\\
M^{10}&=&-\frac{1}{4\pi}[\mu_{t}(2\lambda_{x}+\mu_{x}
-\nu_{x})+\mu_{tx}]~e^{2\lambda-\nu-\mu}.
\end{eqnarray}

\section{Energy-Momentum Distribution in TPT}

In this section, we use the above mentioned four prescriptions in
the context of TPT to evaluate the energy-momentum distribution of
non-static plane symmetric spacetimes. The corresponding tetrad
components of the metric (39) are given as
\begin{equation}
{h^{a}}_{\mu}= \left(
  \begin{array}{cccc}
    e^{\nu(t,x)} & 0 & 0 & 0 \\
    0 & e^{\mu(t,x)} & 0 & 0 \\
    0 & 0 & e^{\lambda(t,x)} & 0 \\
    0 & 0 & 0 & e^{\lambda(t,x)} \\
  \end{array}
\right)
\end{equation}
with its inverse
\begin{equation}
{h_{a}}^{\mu}= \left(
  \begin{array}{cccc}
    e^{-\nu(t,x)} & 0 & 0 & 0 \\
    0 & e^{-\mu(t,x)} & 0 & 0 \\
    0 & 0 & e^{-\lambda(t,x)} & 0 \\
    0 & 0 & 0 & e^{-\lambda(t,x)} \\
  \end{array}
\right).
\end{equation}
One can easily verify Eqs.(2) and (3) with the help of Eqs.(52)
and (53). Substituting  Eqs.(52) and (53) in Eq.(1), we get the
following non-zero components of Weitzenb\"{o}ck connection
\begin{eqnarray}
{\Gamma^{0}}_{00}&=&\nu_{t},\nonumber\\
{\Gamma^{0}}_{01}&=&\nu_{x},\nonumber\\
{\Gamma^{1}}_{10}&=&\mu_{t},\nonumber\\
{\Gamma^{1}}_{11}&=&\mu_{x},\nonumber\\
{\Gamma^{2}}_{20}={\Gamma^{3}}_{30}&=&\lambda_{t},\nonumber\\
{\Gamma^{2}}_{21}={\Gamma^{3}}_{31}&=&\lambda_{x}.
\end{eqnarray}
Eq.(4) then gives the corresponding non-vanishing components of
the torsion tensor as
\begin{eqnarray}
{T^{0}}_{10}=-{T^{0}}_{01}&=&\nu_{t},\nonumber\\
{T^{1}}_{01}=-{T^{1}}_{10}&=&\mu_{t},\nonumber\\
{T^{2}}_{02}=-{T^{2}}_{20}&=&\lambda_{t},\nonumber\\
{T^{2}}_{21}=-{T^{2}}_{12}&=&\lambda_{x}\nonumber\\
{T^{3}}_{03}=-{T^{3}}_{30}&=&\lambda_{t},\nonumber\\
{T^{3}}_{13}=-{T^{3}}_{31}&=&\lambda_{x}.
\end{eqnarray}
Multiplying the above components of the torsion tensor with
relevant $g^{\mu\nu}$ and then using in Eq.(24), we get the
following non-zero components of the tensor $S^{abc}$ as
\begin{eqnarray}
S^{010}=-S^{001}&=& e^{-2(\nu+\mu)}\lambda_{x},\nonumber\\
S^{101}=-S^{110}&=& {e^{-2(\mu+\nu)}} \lambda_t,\nonumber\\
S^{202}=-S^{220}&=& \frac{1}{2}e^{-2(\lambda+\nu)}(\lambda_{t}+\mu_{t}),\nonumber\\
S^{221}=-S^{212}&=& \frac{1}{2}e^{-2(\lambda+\mu)}(\lambda_{x}+\nu_{x}),\nonumber\\
S^{303}=-S^{330}&=& \frac{1}{2}e^{-2(\lambda+\nu)}(\lambda_{t}+\mu_{t}),\nonumber\\
S^{313}=-S^{331}&=&
\frac{1}{2}e^{-2(\lambda+\mu)}(\lambda_{x}+\nu_{x}).
\end{eqnarray}
Making use of Eq.(56) in Eq.(25) yields the following non-zero
components of the superpotential as
\begin{eqnarray}
{U_{0}}^{10}=-{U_{0}}^{01}&=&e^{2\lambda+\nu+\mu}\lambda_{x},\nonumber\\
{U_{1}}^{10}=-{U_{1}}^{01}&=&e^{2\lambda+\mu-\nu}\lambda_{t},\nonumber\\
{U_{2}}^{20}=-{U_{2}}^{02}&=&\frac{1}{2}e^{2\lambda+\mu-\nu}(\lambda_{t}+\mu_{t}),\nonumber\\
{U_{2}}^{12}=-{U_{2}}^{21}&=&\frac{1}{2}e^{2\lambda-\mu+\nu}(\lambda_{x}+\nu_{x}),\nonumber\\
{U_{3}}^{30}=-{U_{3}}^{03}&=&\frac{1}{2}e^{2\lambda+\mu-\nu}(\lambda_{t}+\mu_{t}),\nonumber\\
{U_{3}}^{13}=-{U_{3}}^{31}&=&\frac{1}{2}e^{2\lambda-\mu+\nu}(\lambda_{x}+\nu_{x}).
\end{eqnarray}

Using Eq.(57) in Eqs.(26), (27) and (28), the non-vanishing
energy-momentum density components of Einstein, Landau-Lifshitz
and Bergmann-Thomson prescriptions respectively are
\begin{eqnarray}
hE^{00}&=&-\frac{1}{4\pi}[\lambda_{x}(2\lambda_x-\mu_{x}
+\nu_{x})+\lambda_{xx}]~e^{2\lambda-\nu-\mu},\\
hE^{10}&=&\frac{1}{4\pi}[\lambda_{t}(2\lambda_x+\mu_{x}
-\nu_{x})+\lambda_{tx}]~e^{2\lambda-\nu-\mu},\\
hL^{00}&=&-\frac{1}{4\pi}e^{4\lambda}(4\lambda_{x}^{2}+\lambda_{xx}),\\
hL^{10}&=&\frac{1}{4\pi}e^{4\lambda}(4\lambda_{x}\lambda_{t}+\lambda_{tx})
\end{eqnarray}
and
\begin{eqnarray}
hB^{00}&=&-\frac{1}{4\pi}[\lambda_{x}(2\lambda_{x}-\mu_{x}
-\nu_{x})+\lambda_{xx}]~e^{2\lambda-\nu-\mu},\\
hB^{10}&=&\frac{1}{4\pi}[\lambda_{t}(2\lambda_{x}-\mu_{x}-\nu_{x})
+\lambda_{tx}]~e^{2\lambda-\nu-\mu}.
\end{eqnarray}
Now, using Eq.(55) in Eq.(6) and then multiplying by relevant
components of $g^{\mu\nu}$, we get the following non-vanishing
components of the contorsion tensor in contravariant form as
\begin{eqnarray}
K^{010}=-K^{100}&=&e^{-2(\mu+\nu)}{\nu}_{x},\nonumber\\
K^{101}=-K^{011}&=&e^{-2(\mu+\nu)}{\mu}_{t},\nonumber\\
K^{202}=-K^{022}&=&e^{-2(\lambda+\nu)}{\lambda}_{t},\nonumber\\
K^{122}=-K^{212}&=&e^{-2(\lambda+\mu)}{\lambda}_{x},\nonumber\\
K^{303}=-K^{033}&=&e^{-2(\lambda+\nu)}{\lambda}_{t},\nonumber\\
K^{313}=-K^{133}&=&e^{-2(\lambda+\mu)}{\lambda}_{x}.
\end{eqnarray}
Clearly, the contorsion tensor is antisymmetric w.r.t. its first
two indices. Substituting Eq.(55) in Eq.(35) and then multiplying
by relevant components of $g^{\mu\nu}$ yields the non-zero basic
vector components in contravariant form as
\begin{eqnarray}
V^{0}&=&-e^{-2\nu}(\mu_{t}+2\lambda_{t}),\nonumber\\
V^{1}&=&e^{-2\mu}(\nu_{x}+2\lambda_{x}).
\end{eqnarray}
Substituting Eqs.(39), (64), (65) and $\kappa=8\pi$ ( taking,
$G=c=1$) in Eq.(32), the required non-vanishing components of the
superpotential turn out to be
\begin{eqnarray}
U_0^{01}&=&-\frac{1}{4\pi}\lambda_x~e^{2\lambda-\mu+\nu},\nonumber\\
U_1^{01}&=&-\frac{1}{4\pi}\lambda_t~e^{2\lambda+\mu-\nu}.
\end{eqnarray}
In view of Eq.(66), the non-zero energy-momentum density
components in contravariant form can be obtained from Eq.(36)
after multiplication with $g^{00}$ and $g^{11}$ as
\begin{eqnarray}
\Xi^{00}&=&-\frac{1}{4\pi}[\lambda_{x}(2\lambda_{x}
-\mu_{x}+\nu_{x})+\lambda_{xx}]~e^{2\lambda-\mu-\nu},\nonumber\\
\Xi^{10}&=&\frac{1}{4\pi}[\lambda_{t}(2\lambda_{x}
+\mu_{x}-\nu_{x})+\lambda_{tx}]~e^{2\lambda-\mu-\nu}.
\end{eqnarray}

\section{Summary and Discussion}

Energy-momentum is an important conserved quantity whose
definition has been under investigation since the birth of GR.
Although, the problem of localization of energy is unresolved and
controversial but much attention has been given by different
scientists to resolve it. Here, we have discussed the problem of
localization of energy-momentum in two different frameworks of GR
and TPT by using different energy-momentum complexes.
 We used Einstein, Landau-Lifshitz, Bergmann-Thomson and
M{\o}ller prescriptions to explore the energy-momentum
distribution of non-static plane symmetric spacetimes in the
context of both GR and TPT. Although, on the basis of this work we
are not able to resolve the longstanding and crucial problem of
the localization of energy but it adds one more example which may
be used to make a conjuncture about the localization of energy at
some stage. The results obtained so far are given in the following
tables (1-8):

\vspace{0.5cm}

{\bf {\small Table 1.} {\small Energy-Momentum Density (EMD)
Components of Einstein's Prescription in GR }}
\begin{center}
\begin{tabular}{|c|c|}
\hline{\bf EMD}&{\bf Expressions}\\
\hline{$\Theta^{00}$} &

$-\frac{1}{4\pi}[\lambda_{x}(2\lambda_{x}-
\mu_{x}+\nu_{x})+\lambda_{xx}]~e^{2\lambda-\mu-\nu}$\\

\hline{ $\Theta^{10}$} &

$ \frac{1}{4\pi}[\lambda_{t}(2\lambda_{x}+\mu_{x}-\nu_{x})+
\lambda_{tx}]~e^{2\lambda-\mu-\nu}$\\ \hline
\end{tabular}
\end{center}

\vspace{0.5cm}

{\bf {\small Table 2.} {\small Energy-Momentum Density (EMD)
Components of Einstein's Prescription in TPT }}
\begin{center}
\begin{tabular}{|c|c|}
\hline{\bf EMD}&{\bf Expressions}\\
\hline h$E^{00}$ & $-\frac{1}{4\pi}[\lambda_{x}(2\lambda_{x}-
\mu_{x}+\nu_{x})+\lambda_{xx}]~e^{2\lambda-\mu-\nu}$\\ \hline
h$E^{10}$ & $
\frac{1}{4\pi}[\lambda_{t}(2\lambda_{x}+\mu_{x}-\nu_{x})+\lambda_{tx}]
~e^{2\lambda-\mu-\nu} $\\ \hline
\end{tabular}
\end{center}

\vspace{0.5cm}

{\bf {\small Table 3.} {\small Energy-Momentum Density (EMD)
Components of Landau-Lifshitz's Prescription in GR }}
\begin{center}
\begin{tabular}{|c|c|}
\hline{\bf EMD}&{\bf Expressions}\\
\hline $\L^{00}$ & $
-\frac{1}{4\pi}(4\lambda_{x}^{2}+\lambda_{xx})~e^{4\lambda}
$\\
\hline $\L^{10}$ &  $
\frac{1}{4\pi}(4\lambda_{x}\lambda_{t}+\lambda_{tx})~e^{4\lambda}
$\\
\hline
\end{tabular}
\end{center}

\newpage
\vspace{0.5cm}

{\bf {\small Table 4.} {\small Energy-Momentum Density (EMD)
Components of Landau-Lifshitz's Prescription in TPT}}

\begin{center}
\begin{tabular}{|c|c|}
\hline{\bf EMD}&{\bf Expressions}\\
\hline h$L^{00}$ & $
-\frac{1}{4\pi}(4\lambda_{x}^{2}+\lambda_{xx})~e^{4\lambda}
$\\
\hline h$L^{10}$ & $
\frac{1}{4\pi}(4\lambda_{x}\lambda_{t}+\lambda_{tx})~e^{4\lambda}
$\\
\hline
\end{tabular}
\end{center}

\vspace{0.5cm}

{\bf {\small Table 5.} {\small Energy-Momentum Density (EMD)
Components of Bergmann-Thamson's Prescription in GR }}

\vspace{0.5cm}

\begin{center}
\begin{tabular}{|c|c|}
\hline{\bf EMD}&{\bf Expressions}\\
\hline $B^{00}$ & $-\frac{1}{4\pi}[\lambda_{x}(2\lambda_{x}-
\nu_{x}-\mu_{x})+\lambda_{xx}]~e^{2\lambda-\mu-\nu}$\\
\hline $B^{10}$ & $
\frac{1}{4\pi}[\lambda_{t}(2\lambda_{x}-\mu_{x}-\nu_{x})+\lambda_{tx}]~e^{2\lambda-\mu-\nu}
$\\
\hline
\end{tabular}
\end{center}

\vspace{0.5cm}

{\bf {\small Table 6.} {\small Energy-Momentum  Density (EMD)
Components of Bergmann-Thamson's Prescription in TPT}}

\vspace{0.5cm}

\begin{center}
\begin{tabular}{|c|c|}
\hline{\bf EMD}&{\bf Expressions}\\
\hline h$B^{00}$ & $-\frac{1}{4\pi}[\lambda_{x}(2\lambda_{x}-
\nu_{x}-\mu_{x})+\lambda_{xx}]e^{2\lambda-\mu-\nu}$\\
\hline h$B^{10}$ & $
\frac{1}{4\pi}[\lambda_{t}(2\lambda_{x}-\mu_{x}-\nu_{x})+\lambda_{tx}]e^{2\lambda-\mu-\nu}
$\\
\hline
\end{tabular}
\end{center}

\vspace{0.5cm}

{\bf {\small Table 7.} {\small Energy-Momentum  Density (EMD)
Components of M{\o}ller's Prescription in GR }}

\begin{center}
\begin{tabular}{|c|c|}
\hline{\bf EMD}&{\bf Expressions}\\
\hline $M^{00}$ & $

\frac{1}{4\pi}[\nu_{x}(2\lambda_{x}-\mu_{x}+\nu_{x}
)+\nu_{xx}]~e^{2\lambda-\mu+\nu}
$\\
\hline $M^{10}$ & $

-\frac{1}{4\pi}[\mu_{t}(2\lambda_{x}+\mu_{x}-\nu_{x}
)+\mu_{tx}]~e^{2\lambda-\mu+\nu}
$\\
\hline
\end{tabular}
\end{center}

\vspace{0.5cm}

{\bf {\small Table 8.} {\small Energy-Momentum(E-M) densities in
M{\o}ller's Prescription in TPT}}

\vspace{0.5cm}

\begin{center}
\begin{tabular}{|c|c|}
\hline{\bf EMD}&{\bf Expressions}\\
\hline $\Xi^{00}$ & $ -\frac{1}{4\pi}[\lambda_{x}(2\lambda_{x}
-\mu_{x}+\nu_{x})+\lambda_{xx}]~e^{2\lambda-\mu-\nu}
$\\
\hline $\Xi^{10}$ & $

\frac{1}{4\pi}[\lambda_{t}(2\lambda_{x}+\mu_{x}-
\nu_{x})+\lambda_{tx}]~e^{2\lambda-\mu-\nu}
$\\
\hline
\end{tabular}
\end{center}

\vspace{0.5cm}

These tables show that the energy-momentum density components turn
out to be well defined and finite for each prescription in both GR
and TPT. It is mentioning here that the only non-vanishing
component of the momentum density is along x-axis while the other
components turn out to be zero in each case. It is due to the fact
that we have considered the metric in which the metric functions
are depending on $t$ and $x$ only. Also, we can obtain the
corresponding results along $y$- or $z$-axes by considering the
metric function depending on $y$ or $z$ along with $t$. From the
results given in tables $1-6$, it is noted that the three
prescriptions, namely, Einstein, Landau-Lifshitz and
Bergmann-Thomson (ELLBT) yield the same energy-momentum
distribution of non-static plane symmetric spacetimes in both GR
and TPT, while the results of M{\o}ller's prescriptions in both
the theories are different. Further, tables $2,8$ show that the
energy as well as momentum density components of Einstein's and
Moller's prescriptions turn out to be same in TPT. It is worth
mentioning here that our results coincide with the results
obtained by Sharif and Kanwal [1] for Bell-Szekeres metric under
certain choice of the metric functions.

In the end, it is suggested that the issue of localization of
energy may be tackled in some other theories , like \emph{f(r)}
theory of gravity.

\vspace{0.5cm}


{\bf Acknowledgment}

\vspace{0.5cm}

We acknowledge the enabling role of the Higher Education
Commission Islamabad, Pakistan, and appreciate its financial
support through the {\it Indigenous PhD 5000 Fellowship Program
Batch-III}.

\newpage

\vspace{0.5cm}

{\bf References}

\begin{description}

\item{[1]}  M. Sharif and K. Nazir, Braz. J. Phys. \textbf{38},
            156(2008).

\item{[2]}  C.W. Misner, K.S. Thorne,  and J.A. Wheeler, \textit{Gravitation}
            (Freeman, New York, 1973).

\item{[3]} A. Einstein, Sitzungsber. Preus. Akad. Wiss. Berlin (Math. Phys.)
            778(1915), Addendum ibid 779(1915).

\item{[4]} L.D. Landau,  and E.M. Lifshitz, \textit{The Classical Theory of Fields}
           (Addison-Wesley Press, New York, 1962).

\item{[5]} C. M{\o}ller, Ann. Phys. (N.Y.) \textbf{4}, 347(1958).

\item{[6]} P.G. Bergmann,  and R. Thomson, : Phys. Rev.
\textbf{89}, 400(1958).

\item{[7]} S. Weinberg, \textit{Gravitation and Cosmology} (Wiley, New York, 1972).

\item{[8]} R.C. Tolman, \textit{Relativity, Thermodynamics and Cosmology }
           (Oxford University Press, Oxford, 1934).

\item{[9]} F.I. Cooperstock,  and R.S. Sarracino, J. Phys. A: Math. Gen.
           \textbf{11}, 877(1978).

\item{[10]} K.S. Virbhadra, \textbf{D60}, 104041(1999); \textbf{D42},
2919(1990); Phys. Lett. \textbf{B331}, 302(1994).

\item{[11]} K.S. Virbhadra,  and J.C. Parikh, Phys. Lett.
            \textbf{B317}, 312(1993).

\item{[12]} N. Rosen, and K.S. Virbhadra, Gen. Gravit.  \textbf{25}, 429(1993).

\item{[13]} R. Penrose, Proc. Roy. Soc., London \textbf{A381}, 53(1982).

\item{[14]} K.P. Tod, \textit{Proc. Roy. Soc., London }\textbf{A388}, 457(1983).

\item{[15]} A. Einstein, Sitzungsber. Preuss. Akad. Wiss, 217(1928).

\item{[16]} K. Hayashi,  and T. Nakano, : Prog. Theor. Phys. \textbf{38}, 491(1967).

\item{[17]} R. Weitzenb\"{o}ck, \textit{Invarianten Theorie} (Gronningen: Noordhoft, 1923).

\item{[18]} K. Hayashi,  and T. Shirafuji, : Phys. Rev.
            \textbf{D19}, 3524(1979).

\item{[19]} F.I. Mikhail, M.I. Wanas, A. Hindawi,  and E.I. Lashin, Int. J. Theo.
            Phys. \textbf{32}, 1627(1993).
            F.W. Hehl,  and A. Macias, Int. J. Mod. Phys. \textbf{D8}, 399(1999);
            Obukhov, Yu N., Vlachynsky, E.J., Esser, W., Tresguerres, R. and
            F.W. Hehl,  Phys. Lett. \textbf{A220}, 1(1996);
            P. Baekler, M. Gurses, F.W. Hehl,  and J.D. McCrea, Phys. Lett.
            \textbf{A128}, 245(1988); E.J. Vlachynsky,  W. Esser, W.X. Tresguerres and F.W. Hehl,
             Class. Quantum Grav. \textbf{13}, 3253(1996); J.K. Ho, D.C. Chern and
            J.M. Nester, Chin. J. Phys. \textbf{35}, 460(1997).

\item{[20]} T. Vargas, Gen. Relat. Gravit. \textbf{30}, 1255(2004).

\item{[21]} N. Rosen, Gen. Relat. Gravit. \textbf{26}, 323(1994).

\item{[22]} J.G. Pereira, T. Vargas,  and C.M. Zhang, Class. Quantum Grav.

             \textbf{18}, 833(2001).

\item{[23]} M. Sharif,  and M. J. Amir, Gen. Relat. Gravit.\textbf{ 38}, 1735(2006).

\item{[24]} M. Sharif,  and M. J. Amir, Mod. Phys. Lett. \textbf{A22}, 425(2007).

\item{[25]} M. Sharif,  and M. J. Amir, Gen. Relat. Gravit. \textbf{39}, 989(2007).

\item{[26]} M. Sharif,  and M. J. Amir, Canadian J. Phys. \textbf{86}, 1091(2008).

\item{[27]} M. Sharif,  and M. J. Amir, Mod. Phys. Lett. \textbf{A23}, 3167(2008).

\item{[28]} M. Sharif,  and M. J. Amir, Canadian J. Phys. \textbf{86}, 1297(2008).

\item{[29]} M. Sharif,  and M. J. Amir, Int. J. Theor. Phys. \textbf{47}, 1742(2008).

\item{[30]} M. Sharif,  and S. Taj, Astrophys. Space Sci. \textbf{325}, 75(2010).

\item{[31]} R. Aldrovendi,  and J.G. Pereira, \textit{An Introduction to Gravitation
            Theory}(Preprint).

\item{[32]} A. Trautman, \textit{Gravitation: An Introduction to
            Current Research, ed. Witten, L} (Wiley, New York,
            1962).

\item{[33]} D. Kramer, H. Stephani, E. Hearlt, and M.A.H. MacCallum, \textit{Exact
            Solution of Einstein's Field Equations} (Cambridge University Press, 2003).
\end{description}
\end{document}